%% file: sample-acmsmall.tex
\documentclass[format=acmsmall, review=false, screen=true]{acmart}

\renewcommand\footnotetextcopyrightpermission[1]{} 
\usepackage{booktabs} 
\usepackage[normalem]{ulem}
\usepackage[ruled]{algorithm2e} 

\SetAlFnt{\small}
\SetAlCapFnt{\small}
\SetAlCapNameFnt{\small}
\SetAlCapHSkip{0pt}
\IncMargin{-\parindent}

\begin{document}

\title
{Impact of News Organizations' Trustworthiness and Social Media Activity on Audience Engagement}

\author{Bhavtosh Rath}
\affiliation{%
  \institution{University of Minnesota}
  \country{USA}}
\email{rathx082@umn.edu}
\author{Jisu Kim}
\affiliation{%
  \institution{University of Minnesota}
  \country{USA}
}
\email{kimx4290@umn.edu}
\author{Jisu Huh}
\affiliation{%
 \institution{University of Minnesota}
 \country{USA}}
\email{jhuh@umn.edu}
\author{Jaideep Srivastava}
\affiliation{%
  \institution{University of Minnesota}
  \country{USA}
}
\email{srivasta@umn.edu}

\begin{abstract}
News organizations are increasingly using social media to reach out to their audience aimed at raising their attention and engagement with news. Given the continuous decrease in subscription rates and audience trust in news media, it is imperative for news organizations to understand factors contributing to their relationships with the audience. Using Twitter data of 315 U.S. newspaper organizations and their audiences, this study uses multiple regression analysis to examine the influence of key news organization characteristics on audience engagement with news: (1) trustworthiness computed by the Trust Scores in Social Media (TSM) algorithm; (2) quantity of tweets; and (3) skillfulness of Twitter use. The results show significant influence of a news organizations' trustworthiness and level of Twitter activity on its audiences' news engagement. Methods to measure trustworthiness of news organizations and audience news engagement, as well as scalable algorithms to compute them from large-scale datasets, are also proposed.
\end{abstract}

%
%


%
%

\keywords{Trustworthiness; Social Media Activity; Audience Engagement; Multiple Regression.}

\maketitle


\input{samplebody-journals}

\end{document}

%% file: samplebody-journals.tex
\section{Introduction}
Social media has ubiquitous influence on our daily lives. Since social media is designed to spread and exchange information rapidly through users' interaction, a number of different organizations and brands use social media platform to reach their consumers, stakeholders and audiences for raising awareness and engagement \cite{ashley2015creative, lovejoy2012engaging}.  

News organizations are also increasingly using social media to reach out to their audience aimed at raising audience engagement with news \cite{olmstead2011navigating}. Given the continual fall in subscription rates \cite{goldsmith17} and thus guaranteed engagement, it is imperative for news organizations to find alternative channels to distribute news and directly interact with their audience. Specifically, Twitter has become a popular social media platform for journalists and news organizations to share news with the audience, with the expectation of increasing online traffic to their websites \cite{hermida2014sourcing, hong2012online, messner2011shoveling}. 

According to Messner et al. \cite{messner2011shoveling}, more than 99\% of the top 199 news organizations in the U.S. had created official Twitter accounts by 2010 and regularly post their news with hyperlinks to direct users to the news organizations' websites. When it comes to the audience, more than 70\% of U.S. adults get news from social media and Twitter is the third largest social media platform in terms of the size of news source platforms (after Facebook and Youtube) \cite{shearer17}. More importantly, the audience on social media not only consumes news but also actively engages with the content. For example, people engage with news tweets by liking, sharing (i.e., retweeting) and commenting (i.e., replying) them on Twitter \cite{mitchell16}. These kinds of audience engagement behaviors could help news organizations to increase the visibility of news content, and thus widen their digital presence. The increasing visibility and presence would help news organizations have a more loyal and engaged audience. Hence there is an increasing need for news organizations to understand factors that affect the audiences' engagement behaviors with their content on social media.

This study explores the influence of three important characteristics of news organizations on audience engagement with news: (1) trustworthiness computed by the Trust Scores in Social Media (TSM) algorithm; (2) quantity of tweets; and (3) skillfulness of Twitter use. This paper proposes methods to measure trustworthiness of news organizations, the quality of news organizations' Twitter use, and the extent to which the audience engages with news content by using a large dataset based on Twitter activities of 315 news organizations in the U.S and their audience.  We examine a model that shows how trustworthiness and activities of news organizations on Twitter influence audience engagement behaviors with news. We examine the following three specific Research Questions (\textbf{RQ}):

\begin{description}
\item[RQ 1: ]How does a news organizations' trustworthiness affect the audiences' engagement with its content on social media?
\item[RQ 2: ]How does the quantity of tweets uploaded by a news organization affect the audiences' engagement with its content on social media?
\item[RQ 3: ]How does the skillfulness of Twitter use of a news organization affect  the audiences' engagement with its content on social media?
\end{description}

The remainder of the paper is divided as follows: In the Background section we provide a literature survey of research on factors that influence audiences' engagement, computational trust research and how news organizations use Twitter. In the Methods section we  provide the intuition behind the formulation. Later, we give a brief overview of the data collection process, followed by detailed explanation of different equations designed for analysis in the Experimental Analysis section. In the Discussions section we analyze our findings in the context of the outcome of the experiments followed by summary of findings. We finally conclude the paper by mentioning the limitations of our paper and provide conclusion and scope of future work.
\\
The contributions of our paper are as follows:
\begin{enumerate}
\item We propose new methods based on Twitter data that are used to analyze how a news organizations' trustworthiness and social media activity affect audience engagement.
\item To the best of our knowledge, this is the first proposed approach to compute trustworthiness of news organizations using Computational Trust concept to understand engagement level of users in a social media platform.
\item The experimental framework proposed in the paper can be extended beyond news organizations to other domains that might be interested to know about factors that affect engagement of social media users.
\end{enumerate}

\section{BACKGROUND and RELATED WORK}
\subsection{Audience News Engagement}
Despite their importance, there has been little research on factors that influence the audiences' news engagement, especially related to characteristics of individual news organizations. News engagement has been defined as all types of experiences the audience has with news content, including their attention to news, cognitive, emotional, or behavioral reactions toward news \cite{oh2015clicking}. Particularly, this study focuses on the audiences' behavioral engagement with news by liking, commenting (i.e., replying), and sharing (i.e., retweeting) news on Twitter. The majority of research on antecedent factors for audiences' news engagement has focused on features of content of news articles \cite{arapakis2014user,ziegele2014creates} and individual users' motivations or behaviors \cite{lee2012news,picone2016shares}. This study explores how (1) trustworthiness of news organizations, (2) the quantity of tweets from each news organizations' Twitter account, and (3) the news organizations' skillfulness of Twitter use (e.g., retweeting other users' posts, mentioning other users or using hashtags) as key factors related to the characteristics of news organizations that affect the audiences' engagement behaviors.

\subsection{Trustworthiness of news organizations}
Trust theory has been used in various fields, including mass communication, political science, economics, sociology, and computer science \cite{sabater2005review} as a factor that influences behaviors of actors in all kinds of relationships between human -to- human, human -to- organization, and organization -to-  organization. Trust in an organization has been understood as the trustors' willingness to rely on the organizations' behaviors based on their evaluation of the organization \cite{mayer1995integrative}. Therefore, from the organizations' perspective, trustworthiness refers to the extent to which their partners (e.g., audience) are willing to rely on the organizations' behaviors. According to the dictionary, trustworthiness refers to an ability to be relied on as honest or truthful. When it comes to the audiences' trust in a news organization, scholars have argued that trust would affect their behaviors toward the organization. In terms of reflecting the relationships between a news organization and its audience, audience engagement with news uploaded by the news organization might be regarded as trust-related behaviors, which could be influenced by their trust in the news organization. That is why this study explores influences of trustworthiness of news organization as a characteristic of organization on the audiences' engagement behaviors.

Unlike the other fields which focus on the trustors' beliefs or perceptions, the domain of Computational Trust has extensively borrowed ideas from graph theoretic models like PageRank\cite{page1999pagerank} and HITS\cite{kleinberg1999authoritative} to examine the trustors' behaviors to rank nodes in a network. Researchers have proposed numerous approaches to quantify trust and assign trust scores for nodes in a network \cite{artz2007survey,kamvar2003eigentrust,mishra2011finding,o2004trust}. These approaches  take into consideration various structural aspects depending on the type of network. Trust Score in Social Media algorithm\cite{roy2015computational} is one such algorithm that proposed a pair of complementary scores (trustingness and trustworthiness) to measure trust scores of actors in a social network. Details regarding the algorithm will be discussed in the Methods and Appendix section.

\subsection{News organizations' Twitter use}
In line with Twitters' popularity, journalism scholars have examined how news organizations used Twitter to distribute their news content and interact with the audience \cite{engesser2015frequency,messner2011shoveling}. Although the majority of activities of news organizations on Twitter is related to release and promote their news reports, news organizations also interact with the audience by using technological and connective features of social media. Although there has been no previous research in the journalism field regarding the relationship between news organizations' use of Twitter and audience engagement, there have been accumulated evidence of organizations' strategies of using social media channels to affect users' engagement on the social media platforms \cite{ashley2015creative}. Therefore, it is important to consider not only the quantity, but also the quality of tweets uploaded by news organizations. Based on the conceptualization and operationalization in \cite{engesser2015frequency}, we define the quantity of tweets as the number of all tweets (including retweets) issued by news organizations' account. Quality of tweets is defined by skillfulness of Twitter use. This includes retweeting other users' posts, using hashtag, or mentioning other users using '@' in a tweet. Previous study shows that there were variance in the number of tweets and skillfulness of Twitter use across different news organizations \cite{engesser2015frequency,messner2011shoveling}. That is why these two activities would be regarded as separate characteristics of news organization.

\section{METHODS} 
\subsection{Trust Score in Social Media algorithm}
Trust Score in Social Media (TSM) algorithm \cite{roy2016trustingness} is an iterative matrix convergence algorithm that tries to quantify  level of trust for every user in a social network. The algorithm assigns a pair of complementary scores which are termed as \textit{trustingness} and \textit{trustworthiness} scores. While trustingness is defined as the propensity of an actor to trust his neighbors in the network, trustworthiness is defined as the willingness of the network to trust an individual user. The two concepts can also be mathematically defined. Based on the algorithm, trustworthiness score of a user in a network is defined as sum total of the fraction of weight of an incoming edge to the trustingness of the other user connecting that edge, taking into consideration the involvement score of the network. While trustingness score of a user in a network is defined as sum total of the fraction of weight of outgoing edge to the trustworthiness of the other user connecting that edge, taking into consideration the involvement score of the network. Detailed explanation of TSM algorithm can be found in the Appendix section.


\subsection{Calculating Trustworthiness score}
In each iteration of the TSM algorithm, trustingness and trustworthiness are computed for every node in the network using the following equations:

\begin{align}
ti(v)=&\sum_{\forall x \in out(v)}\left(\frac{w(v,x)}{1+(tw(x))^s}\right) \\
tw(u)=&\sum_{\forall x \in in(u)}\left(\frac{w(x,u)}{1+(ti(x))^s}\right)
\end{align}
\\
where $u$ and $v$ are user nodes, $ti(v)$ and $tw(u)$ are trustingness and trustworthiness scores of $v$ and $u$, respectively, $w(v,x)$ is the weight of edge from $v$ to $x$, $out(v)$ is the set of outgoing edges of $v$, $in(u)$ is the set of incoming edges of $u$, and $s$ is the involvement score of the network. Involvement is basically the potential risk an actor takes when creating a link in the network, which is set to a constant empirically \cite{roy2015computational}.

Once the trust scores are calculated, TSM normalizes the scores by adhering to the normalization constraint so that both the sum of trustworthiness and the sum of trustingness of all nodes in the network equal to 1. More details can be found in \cite{roy2015computational}.

\begin{figure}
  \includegraphics[width=100mm,scale=1]{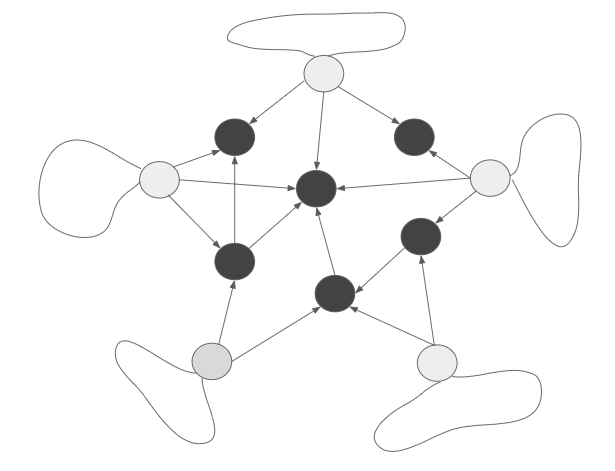}
  \caption{An illustration of the aggregated Twitter network. Grey nodes represent news organizations, black nodes represent their friends and loops around grey nodes represent aggregated follower network.}
  \label{fig:one}
\end{figure}

The TSM algorithm is supposed to include the entire follower-friend network while computing news organizations' trust scores. However news organizations have millions of followers on Twitter (for example The New York Times has over 42 million followers, Washington Post has over 12 million followers, and so on)\footnote{As of February, 2018.}. Thus we used a modified version of TSM. In the initialization stage,  instead of initializing trustingness scores of all news organizations to 1, we initialized the score to trustingness = 1/news organizations' followers count. We then ran the iterative algorithm on network comprising all news accounts and their overlapping friend network. This approach not only helped us to exponentially reduce data collection and algorithm processing time, but also gave good approximations of the news accounts' final trustworthiness scores. The aggregated network used is represented in figure \ref{fig:one}.

\subsection{Social media activity of news organizations}


Regarding activities of news organizations on Twitter, we measured two variables: the quantity of tweets and the skillfulness of Twitter use by each news organization. 

The quantity of tweets measures the number of news   tweets issued by the news organizations' Twitter account for the past two weeks.

\begin{equation}
\label{eqn:3}
Quantity\; of\; tweets=Count\,(news\; tweets)
\end{equation}
\\
The skillfulness of Twitter use measures the quality aspect of Twitter use of news organizations, including (1) tweets that retweets other users' posts or mentions other users in a post (i.e., tweets that refer to other Twitter accounts marked with the at-sign ('@')) and (2) tweets that contain hashtags ('\#'). When a tweet from a news organization includes one of these features ('@' or '\#'), we assigns a score of 1 for the tweet. If a tweet from a news organization includes both features ('@' and '\#'), we assign a score of 2 for the tweet. If the tweet does not include any of these features, we assign a score of 0 for the tweet. These assigned scores would represent the extent to which a tweet includes the \textit{connectivity features}. For each news organization, we calculate the skillfulness of Twitter use score as below.

\begin{equation}
\label{eqn:4}
Skillfulness\; of\; Twitter\; use=\frac{\sum_{\vee\; news\; tweets} Count\,(connectivity\; features)}{Count\,(news\; tweets)}
\end{equation}

\subsection{Engagement score of the audience}
Social media platforms provide different plugin features that help users to engage with the content\cite{almgren2016commenting}. Twitter provides three such plugins: \textit{like}, \textit{retweet} and \textit{reply}. Like conveys a users' positive endorsement of the tweet. Retweet conveys that the user is sharing other users' tweet with his audience. Reply allows a user to comment on other users' tweets. Both retweet and reply actions can be used to convey positive or negative endorsement. An important point to note here is that we excluded retweeted posts while calculating audiences' engagement scores because it was not possible to distinguish audiences' engagement behavior on retweeted posts from audiences' engagement behavior on original posts.

 The following engagement parameters were calculated:
\begin{enumerate}
\item Audience engagement behavior of liking: The sum total of likes for all tweets divided by the total number of tweets.
\item Audience engagement behavior of retweeting: The sum total of retweets for all tweets divided by the total number of tweets.
\item Audience engagement behavior of replying: The sum total of replies for all tweets divided by the total number of tweets.
\end{enumerate}


\section{Data Collection and setup}
We built our dataset from Twitter platform, and all the data used in the experiments was generated from the Twitter API. 315 daily newspaper organizations' official Twitter accounts were selected based on the list from Alliance for Audited Media (accessed on December 29, 2017). We included only English-language newspapers published in the U.S. Puerto Rico Magazines, media groups and trade journal organizations were excluded from our sample. News organizations who did not have activities in this year were also excluded from our sample. 

The network data for calculating the trustworthiness score of news organizations was generated in February, 2018. The variables of social media activity of news organizations and audience engagement were computed based on news tweets collected for a duration of two weeks between March 23, 2018 and April 6, 2018. Data collection stats are shown in table \ref{tab:one} and \ref{tab:two}.
\begin{table}%
\caption{Audience Engagement data overview}
\label{tab:one}
\begin{minipage}{\columnwidth}
\begin{center}
\begin{tabular}{ll}
  \toprule
  Time period of data collection & 3/23/2018 - 4/6/2018 \\
  Average number of likes by audience & 0.877\\
  Average number of retweets by audience & 3.077\\
  Average number of replies by audience & 6.384\\
  \bottomrule
\end{tabular}
\end{center}
\bigskip\centering
\end{minipage}
\end{table}%

\begin{table}%
\caption{News Organization Activity data overview}
\label{tab:two}
\begin{minipage}{\columnwidth}
\begin{center}
\begin{tabular}{ll}
  \toprule
  Number of news organizations     & 315\\
  Total number of news tweets  & 169,867\\
  Total number of news tweets with '@' (mentions)  & 40,546\\
  Total number of news tweets with '\#' (hastags)  & 18,957\\
  \bottomrule
\end{tabular}
\end{center}
\bigskip\centering
\end{minipage}
\end{table}%

\section{Experimental Analysis}
\subsection{Proposed Equations}
We regarded the trustworthiness of news organizations, the quantity of news tweets, and the skillfulness of Twitter use for each news organizations as the independent variables (IVs) and the audience engagement behavior of liking, retweeting and replying as three separate dependent variables (DVs). In addition, since the size of audience for each news organization would affect both the trustworthiness of news organizations and the audiences' engagement behaviors, we need to control this effect. Thus, the circulation of each news organization based on the data from Alliance for Audited Media (accessed on December 29, 2017) was included in the equations as a control variable. 
The following three regression models were used to test our hypothesis:
\\
\begin{equation}
\label{eqn:3}
\textbf{Equation 1}:\; Engagement_1 (Avg.\; number\; of\; likes)=\beta_0 + \beta_1*circulation + \beta_2*tw + \beta_3*QT +\beta_4*STU
\end{equation}
\begin{equation}
\label{eqn:3}
\textbf{Equation 2}:\;Engagement_2 (Avg.\; number\; of\; retweets)=\beta_0 + \beta_1*circulation + \beta_2*tw + \beta_3*QT +\beta_4*STU
\end{equation}
\begin{equation}
\label{eqn:3}
\textbf{Equation 3}:\;Engagement_3 (Avg.\; number\; of\; replies)=\beta_0 + \beta_1*circulation + \beta_2*tw + \beta_3*QT +\beta_4*STU
\end{equation}
\\
where \textit{tw}, \textit{QT} and \textit{STU} stand for trustworthiness score, quantity of tweets and skillfulness of Twitter use respectively.  $\beta$s are the regression co-efficients.

\subsection{Regression Analyses}
In order to examine the proposed equations, we conducted a series of multiple regression analyses (using SPSS package) to analyze how the trustworthiness of news organization, the quantity of tweets, and the skillfulness of Twitter use could predict each audience engagement activity (i.e., liking, retweeting, and replying). A stepwise multiple regression approach was used to find the best combination of independent variables to predict each audience engagement behavior. For the analysis of each dependent variable, the first block entered was the circulation size of news organization using a stepwise method. Next, the trustworthiness score of each news organization was entered, and then in the last block, the quantity of tweets and the skillfulness of Twitter use were entered. The following subsections provide detailed results of the proposed three equations related to each audience engagement behavior.

\subsection{Equation 1 analysis}

With respect to the audience behavior of likes, the first regression model with the circulation size gave the adjusted $R^2$ value of .163, \textit{F}(1, 308) = 61.11, \textit{p} < .001. When we added the trustworthiness scores of news organizations to the second regression model, the adjusted $R^2$ value was .579, \textit{F} (2, 307) = 213.91, \textit{p} < .001. The quantity of tweets (\textit{t} = 0.173, \textit{n.s.}) and skillfulness of Twitter use (\textit{t} = 0.78, \textit{n.s.}) did not enter into the equation at step 3 of the analysis. Overall, the result revealed that the trustworthiness of news organizations emerged as the most prominent predictor of the audience behavior of likes, when the circulation size was controlled for. The final regression equation explained 58\% of the total variance in the dependent variable. Equation 1 results are summarized in table \ref{tab:three}.

\begin{table}%
\centering
\caption{Equation 1 multiple regression stats.}
\label{tab:three}
\begin{minipage}{\columnwidth}
\begin{center}
\begin{tabular}{llll}
  \toprule
  &    \textbf{Independent Variables} & $\textbf{Beta ($\beta$)}^*$\\
    \bottomrule
  \\
 \textbf{Model 1} &Circulation   & .407 \\
 \\
 \multicolumn{3}{l}{\textit{df}=1, 308  \; \textit{F}=61.108  \;\textit{P}=.000 \; $^{**}$\textit{Adjusted $R^2$}=.163} \\
  \midrule
  \\
  \textbf{Model 2} &Circulation  & -.228 \\
 &Trustworthiness   & .905  \\
 \\
  \multicolumn{3}{l}{\textit{df}=2, 307 \; \textit{F}=213.906  \;\textit{P}=.000 \; $^{**}$\textit{Adjusted $R^2$}=.579} \\
  \bottomrule
\end{tabular}
\end{center}
\bigskip\centering
 \textit{ $^{*}$ - p < 0.001,\;($^{*}$, $^{**}$) - for Adjusted $R^2$ indicates significance of $R^2$ increments.}
\end{minipage}
\end{table}%

\subsection{Equation 2 analysis}
Regarding the audience engagement behavior of retweets, the first regression model with the circulation size showed that the variable explained 20.7\% of the variance in audience engagement behavior of retweets, \textit{F}(1, 308) = 81.65, \textit{p} < .001. When trustworthiness was added into the second regression model, the $R^2$ value became .858, indicating that the trustworthiness of news organization accounts for 64.9\% of the variance in the audience engagement behavior of retweets, \textit{F}(2, 307) = 931.50, \textit{p} < .001. Only the quantity of tweets was included into the third regression model, but the $R^2$ change value was .002, \textit{F}(3, 306) = 631.58, \textit{p} < .001. The skillfulness of Twitter use (\textit{t} = 0.48, \textit{n.s.}) did not enter into the equation at step 3 of the analysis. The result showed that the trustworthiness of news organization was still the strongest predictor of the audience engagement behavior of retweets, when the circulation size was controlled for. The final regression equation explained 86\% of the total variance in the dependent variable. Equation 2 results are summarized in table \ref{tab:four}.

\begin{table}%
\centering
\caption{Equation 2 multiple regression stats.}
\label{tab:four}
\begin{minipage}{\columnwidth}
\begin{center}
\begin{tabular}{llll}
  \toprule
  &    \textbf{Independent Variables} & $\textbf{Beta ($\beta$)}^*$\\
    \bottomrule
  \\
 \textbf{Model 1} &Circulation   & .458 \\
 \\
 \multicolumn{3}{l}{\textit{df}=1 \; \textit{F}=81.648  \;\textit{P}=.000 \; $^{**}$\textit{Adjusted $R^2$}=.207} \\
  \midrule
  \\
  \textbf{Model 2} &Circulation  & -.334 \\
 &Trustworthiness   & 1.13  \\
 \\
  \multicolumn{3}{l}{\textit{df}=2 \; \textit{F}=931.5  \;\textit{P}=.000 \; $^{**}$\textit{Adjusted $R^2$}=.858} \\
    \midrule
  \\
  \textbf{Model 3} &Circulation  & -.318 \\
 &Trustworthiness   & 1.145  \\
 &Quantity of tweets   & -.057  \\
 \\
  \multicolumn{3}{l}{\textit{df}=3 \; \textit{F}=631.576  \;\textit{P}=.000 \; $^{**}$\textit{Adjusted $R^2$}=.860} \\
  \bottomrule
\end{tabular}
\end{center}
\bigskip\centering
 \textit{ $^{*}$ - p < 0.001,\;($^{*}$, $^{**}$) - for Adjusted  $R^2$ indicates significance of $R^2$ increments.}
\end{minipage}
\end{table}%

\subsection{Equation 3 analysis}
For the audience engagement behavior of replies, news organizations' circulation size explained 19.6\% of the variance, \textit{F}(1, 308) = 76.23, \textit{p} < .001. The news organizations' trustworthiness explained nearly 65\% of variance in the engagement behavior, \textit{F}(2, 307) = 868.69, \textit{p} < .001. Adding the quantity of tweets in the third model caused the adjusted $R^2$ value to increase to .86, but the $R^2$ change value was .003, \textit{F}(3, 306) = 591.77, \textit{p} < .001. Skillfulness of Twitter use variable is excluded from the third regression model (\textit{t} = 0.20, \textit{n.s.}). Trustworthiness of news organization was the prominent predictor of the audience engagement behavior of replies.
Equation 3 results are summarized in table \ref{tab:five}.

\begin{table}%
\centering
\caption{Equation 3 multiple regression stats.}
\label{tab:five}
\begin{minipage}{\columnwidth}
\begin{center}
\begin{tabular}{llll}
  \toprule
  &    \textbf{Independent Variables} & $\textbf{Beta ($\beta$)}^*$\\
    \bottomrule
  \\
 \textbf{Model 1} &Circulation   & .445 \\
 \\
 \multicolumn{3}{l}{\textit{df}=1 \; \textit{F}=76.23  \;\textit{P}=.000 \; $^{**}$\textit{Adjusted $R^2$}=.196} \\
  \midrule
  \\
  \textbf{Model 2} &Circulation  & -.348 \\
 &Trustworthiness   & 1.132  \\
 \\
  \multicolumn{3}{l}{\textit{df}=2 \; \textit{F}=868.693  \;\textit{P}=.000 \; $^{**}$\textit{Adjusted $R^2$}=.849} \\
    \midrule
  \\
  \textbf{Model 3} &Circulation  & -.330 \\
 &Trustworthiness   & 1.149  \\
 &Quantity of tweets   & -.065  \\
 \\
  \multicolumn{3}{l}{\textit{df}=3 \; \textit{F}=591.796  \;\textit{P}=.000 \; $^{**}$\textit{Adjusted $R^2$}=.852} \\
  \bottomrule

\end{tabular}
\end{center}

\bigskip\centering

\end{minipage}
 \textit{ $^{*}$ - p < 0.001,\;($^{*}$, $^{**}$) - for Adjusted  $R^2$ indicates significance of  $R^2$ increments}

\end{table}%

\section{Discussion}
In this section we analyze our results in the context of the Research Questions proposed earlier.
\begin{description}
\item[RQ 1: ]How does a news organizations' trustworthiness affect the audiences' engagement with its content on social media?
\end{description}
Among all the independent variables proposed in the model, trustworthiness scores accounted for highest variance of all dependent variables. $\beta$ values from experiments show that a higher trustworthiness score positively impacted audience engagement. It could predict 41.7\%, 64.9\% and 65.1\% of variance of engagement behaviors of like, retweet and reply respectively. This shows that network characteristics of news organizations in social media plays an important factor in deciding how audiences engage with them. Trustworthiness is thus a good predictor of how audience engage with news organization on Twitter.

\begin{description}
\item[RQ 2: ]How does the quantity of tweets uploaded by a news organization affect the audiences' engagement with its content on social media?
\end{description}
We had interesting and unexpected observations from our experiment. The stepwise regression model for engagement behavior of like rejected quantity of tweets altogether because it was found to have no significant regression co-efficients (i.e. $\beta$). Regression models for engagement behaviors of retweet and reply had extremely low co-efficient values for the IV and negatively impacted both models. Thus compared to trustworthiness, quantity of tweets posted by news organization is not a good predictor of audience engagement.

\begin{description}
\item[RQ 3: ]How does the skillfulness of Twitter use of a news organization affect  the audience's engagement with its content on social media?
\end{description}
All the three models rejected skillfulness of Twitter use as an IV because of extremely small regression co-efficient values. This was another unexpected observation. We thus conclude that skillfulness of Twitter use is not a good indicator of audience news engagement.

\section{Summary}
Due to the outreach of social media platforms, it is much easier nowadays to access news and engage with the news organization online. Twitter has been widely used by news organizations to build and maintain the relationships with their audiences. Audiences have in turn responded actively by liking, retweeting, and replying on the tweets uploaded by the news organizations. Despite the aforementioned audience trend and the growth of social media as a platform to consume news \cite{olmstead2011navigating}, there have  not been many studies on how different factors influence the audiences' engagement behaviors with news organizations on social media.

To fill the void in the literature and to contribute to developing effective strategies of news organizations to interact with the audience, this study examines which factor influences the audience engagement behaviors. Particularly, we investigated the influences of trustworthiness of news organizations, the quantity of tweets, and the  skillfulness of Twitter use as predictors of audience engagement with news. In order to answer these questions, we used a large volume of data from news organizations' Twitter accounts and proposed methods to measure these variables based on computational approaches. 

We computed the trustworthiness score for news organizations using the TSM algorithm \cite{roy2016trustingness} on their Twitter network. It is an iterative matrix convergence algorithm that assigns a pair of complementary scores (trustingness and trustworthiness) to every node in a network. A large Twitter network dataset was used to assign trustworthiness scores to the news organizations.

Finally we calculated engagement scores for audience based on like, retweet and reply action. The results suggest that trustworthiness of news organization is the best indicator of audience engagement. We also observed that quantity of tweets have a negative impact on engagement behaviors of retweet and reply while skillfulness of Twitter use is a bad predictor of audience engagement.

As methods based on network structure are stronger indicators of audience engagement than methods based on social media activity, we say that characteristics of audience that news organizations interact with are better predictors of engagement than the quantity of news posted or usage of various connectivity features. We conclude that in order to increase audience engagement news organizations must focus on increasing their social media presence and outreach rather than usage of social media platforms' connectivity features. The news organizations' perceived characteristics by the audience is much important than their practices at predicting engagement behavior.

As the study shows that metrics based on sophisticated computational approaches are good predictors of engagement on social media, it would encourage computational social scientists to delve deeper into this research domain.

\section{Limitations}
We included circulation of each news organization based on data from Alliance for Audited Media as the control variable in our regression model. This was done to take into consideration that the trustworthiness scores of new organization and audiences' engagement score will be affected by the size of news organization. 

We excluded retweeted posts while calculating audiences' engagement scores because it was not possible to distinguish audiences' engagement behavior on retweeted posts from audiences' engagement behavior on original posts. 

We ran the TSM algorithm on the aggregated Twitter network where we aggregated the follower network of news organizations based on their follower count. We did so because it was not possible to generate large follower network of news organizations due to Twitter APIs' rate limit restrictions. Details regarding the approach used are explained in Methods section.

\section{Conclusion and future work}
Use of a variety of social media platforms to engage and interact with others is widespread and increasing, making them excellent mechanisms to study social interactions. With precipitous drop in subscription rates and decreasing trust in social media, this study can help news organizations understand factors that increase audience engagement with their content. Other organizational domains (like competing brands) can adopt this setup to understand audience engagement. We hope that future researchers can use this research as a base to identify other organization domains and build better customized metrics and strategies to make their consumers, the public, and stakeholders engage with them. 

As part of future work we would like to use the entire follower network of news organizations while computing trustworthiness scores. We would also like to distinguish audiences' engagement on retweeted and original tweet. It would be interesting to extend our research to news organizations in other countries and perform a comparative analysis of audience engagement. We would also like to identify and design better metrics and build sophisticated models that would be able to predict audiences' engagement levels.

\bibliographystyle{ACM-Reference-Format}
\bibliography{sample-bibliography}

\section{Appendix}
This section contains explanation regarding the Trust Score in Social Media (TSM) algorithm that is used to compute trustworthiness scores for news organizations. The original work on this algorithm was done in 2015 by Atanu Roy \cite{roy2015computational}. Pseudo code is given in Algorithm 1. 
\begin{algorithm}[t]
 \KwData{1. A directed graph G=(V, E) consisting of vertices and edges with or without weights.\newline
 2. Maximum number of permitted iterations, \textit{k} and/or \newline
 3. Difference of scores between two iterations, $\delta$.}
 \KwResult{A set of trust scores: \{(trustingness, trustworthiness)\} for $\forall$ \textit{v} $\in$V. }
Initialize all \textit{v} $\in$V to (1,1). \\
\For{(i = 1; max(max(\textbar $ti_{i}$(v) - $ti_{i-1}$(v)\textbar), max(\textbar $tw_{i}$(v) - $tw_{i-1}$(v)\textbar) \textless $\delta$ or i $\leq$ k; i++)}{

  \For{each node v $\in$V}{
      Update scores of each vertex using scores from last iteration\newline
      $ti^{'}_{i}$(\textit{v}) = $\sum\limits_{\forall out(v)}^{} \frac{w(v,x)}{(1+(tw_{i-1}(x)^s))}$\newline
      out(\textit{v}) = set of all vertices which are destination vertex of all outgoing edges from \textit{v};
  }
 
  \For{each node v $\in$V}{
      Update scores of each vertex using scores from last iteration\newline
      $tw^{'}_{i}$(\textit{v}) = $\sum\limits_{\forall in(v)}^{} \frac{w(x,v)}{(1+(ti_{i-1}(x)^s))}$\newline
      in(\textit{v}) = set of all vertices which are source vertex of all incoming edges to \textit{v}
	}
$ti_i$ = \textit{Normalize}($ti^{'}_{i}$)\newline 
$tw_i$ = \textit{Normalize}($tw^{'}_{i}$)
}
 \caption{Trust in Social Media (TSM) algorithm}
 \label{TSM1 pseudo code}
\end{algorithm}

\begin{figure}
  \includegraphics[width=100mm,scale=1]{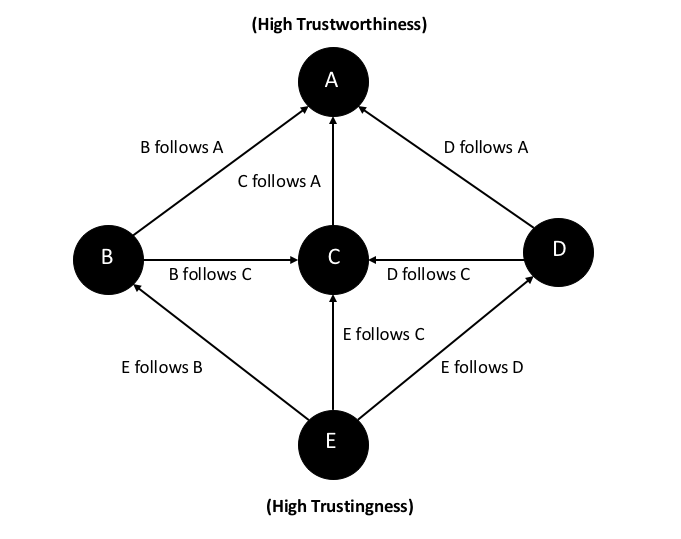}
  \caption{An illustration of trust in Twitter follower network. Nodes represent users and directed edges represent following relationship. Edges indicate source trusting destination.}
  \label{fig:two}
\end{figure}


The primary property leveraged to calculate trust scores is the negative feedback property of trust. The concept of negative feedback in trust can be well understood using
the example Twitter follower network provided in figure~\ref{fig:two}.  There are nodes (say E), which have a high propensity to trust other nodes. E trusts almost all nodes in the network, except 1 (Node A). Thus it can be
seen that E will accord trust to almost anyone in the network which should decrease the weight of its trust vote compared to a node like C which accords its trust very
selectively.
Conversely, it can be seen that node A is a highly trusted node. A high number of nodes in the network trust it. Moreover the nodes that trust it (i.e. B,C,D) in turn trusts a very selective amount of other nodes which makes their (i.e. B,C,D's) votes more valuable compared to E's vote.